\documentclass[twocolumn,showpacs,superscriptaddress]{revtex4}

\usepackage[centertags]{amsmath}
\usepackage{amsfonts}
\usepackage{euscript}
\usepackage{amssymb}
\usepackage{amsthm}
\usepackage{newlfont}
\usepackage{stmaryrd}
\usepackage{mathrsfs}

\newcommand{\ep}{\varepsilon}

\newcommand{\bsP}{\mathbf{P}}

\let\al=\alpha \let\be=\beta \let\de=\delta \let\ep=\epsilon
   
 \let\la=\lambda \let\om=\omega 
\let\si=\sigma   
 \let\up=\upsilon

\let\De=\Delta \let\Ga=\Gamma  \let\Om=\Omega

\newcommand{\caE}{{\mathcal E}}

\newcommand{\caI}{{\mathcal I}}

\newcommand{\opunit}{\text{1}\kern-0.22em\text{l}}

\newcommand{\frI}{{\mathfrak I}}
\newcommand{\frJ}{{\mathfrak J}}

\DeclareMathAlphabet{\mathpzc}{OT1}{pzc}{m}{it}

\newcommand{\cf}{\textit{cf.}}
\newcommand{\eg}{\textit{e.g.}}
\newcommand{\ie}{\textit{i.e.}}

\newcommand{\rel}{\,|\,}

\newcommand{\2}{^{(2)}}

\newcommand{\id}{\textrm{d}}

\begin{document}

\title{Canonical structure of dynamical fluctuations\\
in mesoscopic nonequilibrium steady states}

\author{Christian Maes}
\affiliation{Instituut voor Theoretische Fysica, K.U.Leuven, Belgium}
\email{christian.maes@fys.kuleuven.be}
\author{Karel Neto\v{c}n\'{y}}
\affiliation{Institute of Physics AS CR, Prague, Czech Republic}
\email{netocny@fzu.cz}


\begin{abstract}
We give the explicit structure of the functional governing the
dynamical density and current fluctuations for a mesoscopic system
in a nonequilibrium steady state.  Its canonical form determines a
generalised Onsager-Machlup theory.  We assume that the system is
described as a Markov jump process satisfying a local detailed
balance condition such as typical for stochastic lattice gases and
for chemical networks. We identify the entropy current and the
traffic between the mesoscopic states as extra terms in the
fluctuation functional with respect to the equilibrium dynamics. The
density and current fluctuations are coupled in general, except
close to equilibrium where their decoupling explains the validity of
entropy production principles.
\end{abstract}

\maketitle


Fluctuation theory is at the heart of statistical mechanics. About a
century ago appeared the first fluctuation formul{\ae} regarding
equilibrium systems, in particular from Boltzmann's statistical
interpretation of the thermodynamic entropy. As example, the
equilibrium density fluctuations of a gas in large volume $V$ at
inverse temperature $\beta$ and chemical potential $\mu$ satisfy the
asymptotic law
\begin{equation}\label{eq: einstein}
  \bsP\Bigl(\frac{N}{V} \simeq n\Bigr) \sim
  e^{-\be V\, [\Omega(\mu,n) - \Omega(\mu)]}
\end{equation}
where $\Om(\mu)$ is the grand potential and
$\Om(\mu,n) = F(n) - \mu\, n$ with $F(n)$ the free energy,
is the corresponding variational functional; at least away from the
phase coexistence regime where droplet formation or nucleation
mechanisms become responsible for a slower, surface-exponential
decay. Yet in all cases, there appears an important relation between
the structure of equilibrium fluctuations and the thermodynamics of
the system, making the equilibrium domain exceptionally well
understood. In particular, the variational principles characterising
equilibrium can be understood as an immediate consequence of its
fluctuation theory and response relations can be derived from
expanding \eqref{eq: einstein} around the equilibrium density $n_0$.

In order to include dynamics in the fluctuation theory, Onsager and
Machlup derived the generic structure of small time-dependent
equilibrium fluctuations and explained how their dynamics relates to
the return to equilibrium, \cite{ons}.
The ensuing linear response theory formalised the general relation
between equilibrium current fluctuations and the response in driven
systems in a first-order perturbation theory around equilibrium. To
go beyond and challenged by \eg~the fast progress in nonequilibrium
experiments on nanoscale, one soon realises a lack of general
principles.  Moreover it would be too optimistic to think
nonequilibrium physics based solely on quantities typical to
equilibrium descriptions supplemented with the corresponding
currents.  Deeply related to that is the lack of generally valid
variational principles for nonequilibrium steady states, beyond the
approximate ones of minimum/maximum entropy production.  Yet, more
recently there has also been great progress. One well-known approach
to dynamical (and especially current) fluctuations in open systems
adds to the models fields representing the various reservoirs that
count the long-time statistics of associated `charges' by Master
equation or stochastic path methods, see \eg~\cite{naz,tom,geneve}.
The hydrodynamic fluctuations for some stochastic lattice gas models
have been studied in \eg~\cite{jona,bd}.  For some standard lattice
gas models the large deviations can in fact be explicitly
calculated, see the review~\cite{der}.
Up to now, special emphasis was put on the fluctuations of the
current, also because of relations with a celebrated fluctuation
symmetry of the entropy production, \cite{ecv, GC}.

In the present letter we come back to the basic question whether
there is at all any systematics in the fluctuations beyond
equilibrium or close-to-equilibrium. Can one develop a formalism
that would---similarly to the equilibrium scheme---establish a
link between the dynamical fluctuations and mean
(thermo-)dynamical properties of a system, possibly with the
entropy production playing a role similar to the entropy at
equilibrium? And could that also explain the appearance and
limitations of the entropy-production variational principles on a
fluctuation basis? As we have shown before, \cite{mep}, the
minimum entropy production principle close to equilibrium follows
from the fluctuation theory for the occupation (or residence)
times, which are the relative times spent at different states of
the system. This supports both the relevance of dynamical
fluctuation theory for understanding the status and the validity
of various nonequilibrium variational principles, and the
importance of time-symmetric observables in these considerations.
Indeed our results here strongly suggest that only by treating
jointly the time-symmetric and -antisymmetric sectors do
model-specific results make place for a unique fluctuation
structure.  As far as we know, that is one of the rare occasions
where the nonequilibrium world can be seen submitted to general
laws.

To address the above questions and in the context of a stochastic
network we propose to study the joint dynamical fluctuations of the
occupation times (time-symmetric) and currents (time-antisymmetric).
We will show that these joint fluctuations have an explicit and
general structure, with a fluctuation functional derived from the
so-called traffic measuring the mean dynamical activity in the
system, which is hence the counterpart of the entropy or grand
potential in the equilibrium static fluctuation theory,
\cf~\eqref{eq: einstein}. Only close to equilibrium there emerges a
simple relation between that traffic and the entropy production.
Together with a decoupling between small time-symmetric and
time-antisymmetric fluctuations in the close-to-equilibrium domain,
this lies behind the approximate validity of the entropy production
principles. This substantially extends the argument in~\cite{mep}.
Our main results are relations~\eqref{eq: I-pj1}, \eqref{eq: main},
and \eqref{eq: small} below.  The application of our formalism but
for driven diffusions can be found in~\cite{MNW}.  Our emphasis here
on jump processes makes the analysis also suitable to the statistics
of quantum transport as \eg~in~\cite{fuji,tom}; particular examples
will follow separately.

In contrast with full counting statistics methods, the reservoirs
are not made explicit in our approach. Instead, we assume that the
changes in all reservoirs or loads are mutually distinguishable and
can be read off from the trajectory of the system. Another remark
concerns the meaning of the occupation times. This is a dynamical
observable and its fluctuations fundamentally differ from static
fluctuations, which count the plausibility that the system obeys
some statistics given that the system was in its typical stationary
state far in the past. These static fluctuation functionals are
nonequilibrium variants of the equilibrium free energy and have been
extensively studied in the context of lattice gases in the
hydrodynamic limit, \cite{jona1}. As a matter of principle, one
expects that one could recover the static from the dynamical
fluctuations that are studied here.

The mathematics involved is the theory of large deviations and stems
from the work of Donsker and Varadhan, \cite{DV}. A useful and
repeatedly exploited technique in this approach is to compute the
fluctuation functionals on more coarse-grained levels from
constrained minimisations of a fine-grained functional. That is
called the contraction principle.

\section{General formalism}

We consider a mesoscopic nonequilibrium system modelled by a
stationary ergodic process
$X_t$ making jumps on a discrete set of states, $\{x,y,\ldots\}$. As
is typical for a thermodynamic formalism it is not essential for the
mathematical structure whether the process represents a single
particle random walk or a many-body open system. We only ask the
dynamics as given by some transition rates $w(x,y)$ on ordered pairs
$x \rightarrow y$ to be ergodic.
For an easy interpretation we assume that the local detailed balance
principle applies, according to which $\log\, [w(x,y)/w(y,x)]$ is
the entropy change in the environment (possibly made of several
distinct reservoirs) per single event $x \rightarrow y$. Tracing the
whole trajectory of the system, all currents as well as the total
entropy exchange with the environment can be determined.

We start from that fine-grained level of description and we consider
as dynamical observables the occupation times
\begin{align}
  p_T(x) &= \frac{1}{T} \int_0^T \chi(\om_t = x)\,\id t
\\ \intertext{(with $\chi$ equal to one or zero, indicating whether the event in brackets
 occurs, respectively does not occur)
   jointly with the two-point correlations, for all $x \neq y$,}
  C_T(x,y)\,\de t &= \frac {1}{T}\,
  \int_0^T \chi(\om_t = x)\,\chi(\om_{t+\de t} = y) \, \id t
\end{align}
counting the number of jumps $x
\rightarrow y$, both defined for each realisation of the process
$(\om_t;\,0 \leq t \leq T)$. The occupation times $p_T(x)$ form a
random distribution that asymptotically approaches the stationary
distribution,
$\lim_{T \to \infty} p_T(x) = \rho(x)$, with probability one by the
ergodic theorem. Similarly, the empirical correlations have the
almost-sure asymptotics
$\lim_{T \to \infty} C_T(x,y) = \rho(x) w(x,y)$.

The question about dynamical fluctuations concerns the long-time
asymptotics of possible deviations of $p_T$ and $C_T$ from their
typical values: to compute the probability $\bsP_T(p,k)$ to
observe for all $x$ and $y$,
\begin{equation}\label{eq: constraint}
  p_T(x) = p(x),\qquad C_T(x,y) = p(x) k(x,y)
\end{equation}
We must add here the stationarity condition $\sum_y [p(y) k(y,x) -
p(x) k(x,y)] = 0$ since, by conservation of probability, $\lim_{T
\to \infty} \sum_y [C_T(x,y) - C_T(y,x)] = 0$ for every
realisation of the process.  To determine $\bsP_T(p,k)$ of
\eqref{eq: constraint}, we compare the path-distribution of the
original stationary process with rates $w$ to a fictitious
stationary process with rates $k$. The former distribution reads,
with $\la(x) = \sum_y w(x,y)$ the escape rates,
\begin{multline}
  \bsP_T(\om) = \rho(x_0)\,e^{-\la(x_0)t_1} w(x_0,x_1)\,\id t_1
  \,e^{-\la(x_1)(t_2 - t_1)} \ldots
\\
  \ldots w(x_{n-1},x_n)\,\id t_n\,e^{-\la(x_n)(T - t_n)}
\end{multline}
on realisations $\om = (x_0,0; x_1,t_1; \ldots; x_n,t_n \leq T)$
with jumps $x_{k-1} \rightarrow x_k$ at times $t_k$. The probability
$\bsP^*_T(\om)$ of the same realisation under the fictitious process
is obtained by replacing $w$ with $k$, and $\la$ with the escape
rates $\sum_y k(x,y)$. We exploit that (i) for any trajectory
$\om$ satisfying the constraints~\eqref{eq: constraint} the density of $\bsP_T$ with
respect to the $\bsP_T^*$ only depends on the time-averages $p(x)$
and $k(x,y)$; (ii) those values $p$ and $k$ become
\emph{typical} under the fictitious process $\bsP_T^*$ when $T \to \infty$.
Using both properties, the probability under study is
\begin{equation}
\begin{split}
  \bsP_T(p,k) &= \bsP_T^*(p,k)\,
  \Bigl\langle \frac{\id \bsP_T}{\id \bsP^*_T}
  \,\Bigl|\, \text{conditions }\eqref{eq: constraint}
  \Bigr\rangle_{\bsP_T^*}
\\
  &\sim \frac{\id \bsP_T}{\id \bsP^*_T}(p,k)
\end{split}
\end{equation}
asymptotically for $T \to \infty$. Explicitly,
$\bsP_T(p,k) \sim \exp[-T\,\caI(p,k)]$
with the fluctuation functional
\begin{equation}\label{eq: I-pk}
  \caI(p,k) =
  \sum_{x,y} p(x)\Bigl[k(x,y)\log\frac{k(x,y)}{w(x,y)} - k(x,y) + w(x,y)\Bigr]
\end{equation}
(remember that $I(p,k) = \infty$ whenever $p$ is not stationary with
respect to the transition rates $k$). This result is our starting
point towards a systematic generation of various other fluctuation
laws by contraction, in both the time-symmetric and the
time-antisymmetric domains.

\section{Occupation-current fluctuations}

The observed time-averaged currents correspond to the antisymmetric
part of the two-point correlations, $C_T(x,y) - C_T(y,x)$. The joint
fluctuation law for the currents and the occupation times
\begin{equation}
  \bsP_T(p,j) \sim e^{-T\,I(p,j)}
\end{equation}
can be derived from~\eqref{eq: I-pk} by solving the minimisation
problem
\begin{equation}
  I(p,j) = \inf_k \bigl\{\caI(p,k) \rel
  p(x) k(x,y) - p(y) k(y,x) = j(x,y) \bigr\}
\end{equation}
For stationary currents,
$\sum_y j(x,y) = 0$, to which we from now on solely restrict, the solution
$I(p,j) = \caI(p,k^*)$ is determined from
\begin{align}\nonumber
  k^*(x,y) &= w(x,y)\,e^{\De(x,y)/2}
\\\label{eq: var}
  \De(x,y) &= -\De(y,x)
\\\nonumber
  j(x,y) &= p(x) k^*(x,y) - p(y) k^*(y,x)
\end{align}
(Otherwise $I(p,j) = \infty$.) As a result,
\begin{equation} \label{eq: I-pj1}
  I(p,j) = \frac{1}{4}\sum_{x,y} \De(x,y) j(x,y)
  - \frac{1}{2}\sum_{x,y} [t^*_p(x,y) - t_p(x,y)]
\end{equation}
in which
\begin{align}
  t_p(x,y) = p(x) w(x,y) + p(y) w(y,x)
\\ \intertext{and}
  t^*_p(x,y)= p(x) k^*(x,y) + p(y) k^*(y,x)
\end{align}
measure the mean dynamical activities; we call them
\emph{traffic} and they yield the symmetric counterpart to the
expected currents. The second term in
\eqref{eq: I-pj1} is therefore an \emph{excess} in the overall traffic
needed to create the fluctuation or to make it typical. Similarly,
by the local detailed balance principle, the first term
corresponds to an excess in the entropy flow to the environment
which amounts to
$\dot S = \frac{1}{2}\sum_{x,y} j(x,y) \log [w(x,y) / w(y,x)]$
under the original process and analogously for the modified one.

Next, being motivated by the equilibrium fluctuation theory,
\cf~\eqref{eq: einstein}, we reveal a hidden canonical structure
that enables a particularly illuminating formulation of our result.
Any nonequilibrium process can be related to a reference detailed
balanced one with rates $w_0(x,y)$, so that $w(x,y) =
w_0(x,y)\,e^{\si(x,y) / 2}$ with some driving $\si(x,y) =
-\si(y,x)$. (For example, the rates $w_0(x,y) =
\sqrt{w(x,y)\,w(y,x)}\,e^{s(y) - s(x)}$ and $s$ an arbitrary state
function, can serve as such a reference.) Having fixed $w_0$, the
rates $w(x,y) = w_\si(x,y)$ are now parameterised by the driving
$\si(x,y)$, and we introduce the potential function
\begin{equation}\label{eq: potential0}
  H(p,\si) = 2\sum_{x,y} p(x)\,[w_\si(x,y) - w_0(x,y)]
\end{equation}
equal to the excess in the overall traffic with respect to that
reference. It is a potential for the expected transient currents
$j_{p,\sigma}(x,y) = p(x) w_\sigma(x,y) - p(y) w_\sigma(y,x)$
in the sense that
\begin{equation}\label{eq: potential}
  \de H(p,\si) = \frac{1}{2} \sum_{x,y} j_{p,\si}(x,y)\,
  \de\si(x,y)
\end{equation}
(with the $p$ kept fixed in the variation). Its Legendre transform
is
\begin{equation}\label{eq: G-legendre}
  G(p,j) = \sup_{\si'} \Bigl[ \frac{1}{2} \sum_{x,y} \si'(x,y) j(x,y)
  - H(p,\si') \Bigr]
\end{equation}
and we observe that the supremum (taken over all antisymmetric
matrices) is attained at $\si' = \si^*$ such that $j_{p,\si^*} = j$,
which means that the driving $\si$ and the current $j$ are
canonically conjugated variables. Further, eqs.~\eqref{eq: var} are
solved with
$\De = \si^* - \si$, hence the fluctuation functional
$I(p,j) = I_\si(p,j)$, eq.~\eqref{eq: I-pj1}, obtains the final form
\begin{equation}\label{eq: main}
  I_\si(p,j) = \frac{1}{2} \bigl[ G(p,j) + H(p,\si) - \dot S(\si,j) \bigr]
\end{equation}
with
\begin{equation}\label{eq: ent-current}
  \dot S(\si,j) = \frac{1}{2}\sum_{x,y} \si(x,y) j(x,y)
\end{equation}
the observed entropy current into the environment. That is our main
result, giving the fluctuation functional entirely in terms of the
potential function $H(p,\si)$ (\ie~in terms of the overall traffic)
and derived quantities. The functional $G(p,j)$ directly gives the
reference equilibrium dynamical fluctuations as
$I_0(p,j) = \frac{1}{2} G(p,j)$, hence~\eqref{eq: main} specifies
the nonequilibrium correction to that equilibrium. Remark also that
the antisymmetric part of the functional $I_\si$ under time reversal
equals $I_\si(p,-j) - I_\si(p,j) = \dot S(\si,j)$,
compare~\cite{GC,LS,mn}, which is just the steady state fluctuation
symmetry. However, more important is that~\eqref{eq: main} also in a
generic way specifies the time-symmetric component.  That is
why~\eqref{eq: main} represents a generalised Onsager-Machlup
Lagrangian describing steady fluctuations, the generalised
dissipation functions being $G$ and $H$.  At the same time, one
recognises the mathematical structure of equilibrium fluctuations;
the grand potential $\Om(\mu)$ and the variational functional
$\Om(\mu,n)$ of~\eqref{eq: einstein} get replaced here by
$-H(p,\si) / 2$ and $[G(p,j) - \dot S(\si,j)] / 2$, respectively.
Note that while the potentials $G$ and $H$ depend on the choice of
reference equilibrium dynamics, the resulting functional $I(p,j)$ is
of course independent of that.

Fluctuation laws on a more-coarse grained level, \eg, the
fluctuations of a single selected current, can be obtained by
further contractions starting from~\eqref{eq: I-pk} or~\eqref{eq:
main}. Then, depending on the particular question, a modified
canonical formalism can be established.

There is a trivial yet important generalisation of the above results
to systems in which a transition $x \rightarrow y$ can go via
multiple channels, each possibly corresponding to the interaction
with different reservoirs. For these systems the formulas~\eqref{eq:
I-pk}, \eqref{eq: I-pj1}, \eqref{eq: potential}, \eqref{eq:
G-legendre} etc remain valid if the ordered pairs
$x,y$ in the sums get replaced with $x,y,\al$, the $\al$ labelling
the channels. A simple example of such a multi-channel model comes
in the next section.

\section{Example}

We demonstrate the above formalism on a model of transport over a
single level (quantum dot). There are two configurations
$x = 0,1$ corresponding to the level being empty respectively occupied,
and it is coupled to the left (L) and the right (R) reservoirs.
Using the notation $V_L$ and $V_R$ for the potential gradients
between that level and the reservoirs, both oriented in the
$L\rightarrow R$ direction, the local detailed balance principle restricts the
possible transition rates corresponding to each channel to the
following general form:
\begin{equation}\label{eq: example}
\begin{array}{ll}
  w_L(0,1) = \Ga_L\, e^{\be V_L / 2}, & w_L(1,0) = \Ga_L\, e^{-\be V_L / 2} \\
  w_R(0,1) = \Ga_R\, e^{-\be V_R / 2}, & w_R(1,0) = \Ga_R\, e^{\be V_R / 2}
\end{array}
\end{equation}
For simplicity, we consider here only the case
$\Ga_L = \Ga_R = \Ga$. Writing the occupation times as
$p(0) = (1 - \up)/2$ and $p(1) = (1 + \up)/2$, the
expected transient currents (also both oriented in the $L\rightarrow
R$ direction) and traffic separately for each channel equal
\begin{align}
  j^{L,R}_\up &= \Ga \sinh \frac{\be V_{L,R}}{2} \mp \Ga\up \cosh\frac{\be V_{L,R}}{2}
\\
  t^{L,R}_\up &= \Ga \cosh \frac{\be V_{L,R}}{2} \mp \Ga\up \sinh\frac{\be V_{L,R}}{2}
\end{align}
As a reference equilibrium we take the dynamics~\eqref{eq: example}
for $V_L = V_R = 0$ (with the symmetric part $\Ga$ kept unchanged).
The current potential~\eqref{eq: potential0} is determined from the
overall traffic:
\begin{multline}
  H(\up,V_L,V_R) = 2 \Ga\, \Bigl( \cosh\frac{\be V_L}{2}
  - \up\,\sinh\frac{\be V_L}{2}
\\
  + \cosh\frac{\be V_R}{2}
  + \up\,\sinh\frac{\be V_R}{2} - 2 \Bigr)
\end{multline}
One checks that
$\partial H / \partial V_{L,R} = \be j^{L,R}_\up$ which is an instance
of~\eqref{eq: potential}. The Legendre transform of $H$ at
$j^L = j^R = j$ gives the occupation-current fluctuation functional
$G(\up,j) = I_0(\up,j) / 2$ for the reference equilibrium dynamics,
\cf~\eqref{eq: G-legendre}:
\begin{equation}
\begin{split}
  G(\up,j) &= \sup_{V_L, V_R} \bigl[ \be j(V_L + V_R) - H(\up,V_L,V_R)
  \bigr]
\\
  &= 4 j \log \Bigl[ \frac{1}{\sqrt{1 - \up^2}} \Bigl( \frac{j}{\Gamma}
  + \sqrt{1 - \up^2 + \frac{j^2}{\Gamma^2}} \Bigr) \Bigr]
\\
  &\hspace{5mm}+ 4 \Gamma \Bigl[ 1 - \sqrt{1 - \up^2 + \frac{j^2}{\Gamma^2}} \Bigr]
\end{split}
\end{equation}
This extends to the nonequilibrium dynamics by the generalised
Onsager-Machlup formula~\eqref{eq: main}.  \textit{E.g.}, in the L-R
symmetric case $V_L = V_R = V$, the entropy flow is
$\dot S = 2\be V j$, and the nonequilibrium fluctuation functional
becomes
\begin{multline}
  I_V(\up,j) = 2 j \log \Bigl[ \frac{1}{\sqrt{1 - \up^2}} \Bigl( \frac{j}{\Gamma}
  + \sqrt{1 - \up^2 + \frac{j^2}{\Gamma^2}} \,\Bigl) \Bigr]
  - \be V j
\\
  + 2 \Gamma \Bigl[ \cosh\frac{\be V}{2} -
  \sqrt{1 - \up^2 + \frac{j^2}{\Gamma^2}} \,\Bigr]
\end{multline}
Due to the `particle-hole' symmetry $I(-\up,j) = I(\up,j)$, the
(marginal) current fluctuations correspond to the rate
$\frI_V(j) = I_V(0,j)$, which is
\begin{multline}
  \frI_V(j) = 2 j \log \Bigl[ \frac{j}{\Gamma}
  + \sqrt{1 + \frac{j^2}{\Gamma^2}}\, \Bigr]
  - \be V j
\\
  + 2 \Gamma \Bigl[ \cosh\frac{\be V}{2} -
  \sqrt{1 + \frac{j^2}{\Gamma^2}}\, \Bigr]
\end{multline}
Again by contraction, the fluctuation functional for the occupation
times is
$\frJ_V(\up) = I_V(\up,j^*)$ where
$j^* = \Gamma \sqrt{1 - \up^2}\,\sinh(\be V / 2)$ is the most probable value of the stationary
current given $\up$. As a result,
\begin{equation}
  \frJ_V(\up) = 2 \Gamma \cosh\Bigl(\frac{\be V}{2}\Bigr)\, (1 - \sqrt{1 - \up^2})
\end{equation}

\section{Regime of small fluctuations}

The main features of the joint occupation-current fluctuations
already become manifest in the leading order around the
nonequilibrium steady state. For our original dynamics with
stationary distribution
$\rho$, steady current $\bar j$ and steady traffic $\bar t$,  we
write $p = \rho(1 + \ep\, u_1)$, $j =
\bar j + \ep j_1$.  Standard perturbation theory applied
to~\eqref{eq: I-pj1}, up to quadratic order in $\ep$, gives as a
final result $I(p,j) = \ep^2 I\2(u_1,j_1)$ where
\begin{multline}\label{eq: small}
  I\2(u_1,j_1) = \frac{1}{4} \sum_{x,y} \Bigl[
  \frac{1}{2\bar t}\, j_1^2 + \frac{\bar t}{2}\, (\nabla^- u_1)^2
\\
  - \frac{\bar j}{\bar t}\, j_1 \nabla^+ u_1
  + \frac{\bar j^2}{2 \bar t}\, (\nabla^+ u_1)^2
  \Bigr](x,y)
\end{multline}
with the shorthand $\nabla^\pm u_1(x,y) = [u_1(x) \pm u_1(y)]/2$.
This formula clearly demonstrates how the occupation times and
current become coupled away from equilibrium. That coupling is
proportional to the stationary current and vanishes only in the
close-to-equilibrium regime where $\bar j = O(\ep)$.

The appearance/disappearance of the occupation-current correlation
is deeply related with the validity/breaking of the entropy
production principles. The expected value of the (transient) entropy
production rate $\caE(p)$ at a given distribution $p$ is the sum of
the expected entropy flow
$\frac{1}{2} \sum_{x,y} j_p(x,y) \log\,[w(x,y) / w(y,x)]$ and the rate
of increase of the system's entropy
$-\frac{1}{2} \sum_{x,y} j_p(x,y) \log p(x)$, \cite{LS}.
In the same quadratic approximation as above but now close to
equilibrium so that $w = w_0[1 +O(\ep)]$, the entropy production
rate equals
\begin{equation}
  \caE(p) = \sum_{x,y} \Bigl[ \frac{\ep^2 \bar t}{2}\, (\nabla^- u_1)^2
  + \frac{\bar j^2}{2\bar t} \Bigr](x,y)
\end{equation}
with $\bar j = O(\ep)$.  On the other hand, from~\eqref{eq: small}
the marginal distribution of the occupation times for $\bar j =
O(\ep)$ corresponds to the functional
$\frJ\2(u_1) = \frac{1}{8} \sum_{x,y} \bar t (\nabla^- u_1)^2(x,y)$, and hence
\begin{equation}
  \frJ(p) = \frac{1}{4} \bigl[ \caE(p) - \caE(\rho) \bigr]
\end{equation}
see~\cite{mep} for more details. Hence, the stationary distribution
$\rho$ is a minimiser of the entropy production rate and the
latter governs the occupation fluctuations---this is no longer true
beyond the close-to-equilibrium regime where the occupation-current
correlation becomes relevant. A similar argument reveals a direct
link between the current fluctuations and the maximum entropy
production principle, \cite{MNW}.

\section{Conclusions and remarks}

We have derived an explicit formula for the functional governing the
joint dynamical fluctuations of transition intensities and
occupation times in a steady state regime described by a Markov jump
process, \eqref{eq: I-pk}. In the occupation-current form~\eqref{eq:
main}, it gets a remarkable canonical structure: the (reference)
equilibrium functional is corrected by its Legendre transform which
is just a potential for the expected currents, and by the entropy
flow. These functionals form a natural starting point towards the
study of fluctuations for any selected collection of observables
that can be expressed in terms of transitions/currents and
occupations, via the contraction principle. That provides an
alternative to the existing approaches.

As a new and crucial quantity, unseen in close-to-equilibrium
considerations, enters the traffic, measuring the time-symmetric
dynamical activity in the system. This observable naturally enters
beyond the linear response theory, \eg, in determining the ratchet
current~\cite{woj} and in the escape rate theory, \cite{hang}. The
overall traffic yields the current potential, and its excess
together with an excess in the entropy flow directly determine the
joint occupation-current fluctuations,
\eqref{eq: I-pj1}.

The time-symmetric and time-antisymmetric fluctuations mutually
couple even for small fluctuations around the nonequilibrium state,
\eqref{eq: small}. Their decoupling in leading order around equilibrium is a
fundamental reason for the known stationary variational principles
to be approximately valid.

For extended systems with a large number of degrees of freedom,
phase transitions may become visible through singularities of the
fluctuation functionals, \cite{trans}. It should indeed not escape
the attention that the analysis from~\eqref{eq: I-pj1}
to~\eqref{eq: main} requires some strict convexity arguments and
uniqueness of solutions.  That is certainly one of the most
fascinating possibilities that can be discussed within our general
framework.


\begin{acknowledgments}
K.N.\ is grateful to Tom\'a\v{s} Novotn\'y for fruitful discussions
and suggestions, and also acknowledges the support from the Grant
Agency of the Czech Republic (Grant no.~202/07/J051). C.M.\ benefits
from the Belgian Interuniversity Attraction Poles Programme P6/02.
\end{acknowledgments}


\end{document}